\documentclass[a4paper,11pt]{article}
\usepackage{pos}
\usepackage{amsmath,latexsym,bbm}
\usepackage{braket}
\usepackage{bm}
\usepackage{booktabs}
\usepackage{setspace}
\usepackage{hepunits}
\usepackage{graphicx}% Include figure files

\usepackage{mciteplus}

\newcommand{\nn}{\nonumber}
\newcommand{\nd}{\mathrm{d}}
\newcommand{\nD}{\mathrm{D}}
\newcommand{\nT}{\mathrm{T}}
\newcommand{\nW}{\mathrm{W}}
\newcommand{\nds}{\hat{\mathrm{d}}}

\newcommand{\ep}{\epsilon}

\allowdisplaybreaks

\DeclareMathOperator{\Res}{Res}

\title{Selection rules of canonical differential equations from Intersection theory}
\author*[a]{Jiaqi Chen}
\affiliation[a]{Beijing Computational Science Research Center,\\
  Beijing 100084, China}

\emailAdd{jiaqichen@csrc.ac.cn}
\abstract{The matrix of canonical differential equations consists of the 1-$\mathrm{d}\log$-form coefficients obtained by projecting ($n$+1)-$\mathrm{d}\log$-forms onto $n$-$\mathrm{d}\log$-form master integrands. With dual form in relative cohomology, the intersection number can be used to achieve the projection and provide the selection rules for canonical differential equations, which relate to the pole structure of the $\mathrm{d}\log$ master integrands.}

\FullConference{Loops and Legs in Quantum Field Theory (LL2024)\\
 14-19, April, 2024\\
Wittenberg, Germany\\}

\begin{document}
\maketitle

\section{Introduction}
 Integration-By-Parts (IBP) \cite{Chetyrkin:1981qh} and IBP-based differential equations \cite{Kotikov:1990kg,Kotikov:1991pm,Gehrmann:1999as,Bern:1993kr} are two of the most powerful methods in perturbative quantum field theory. By appropriately selecting the master integrals, their differential equations can be transformed into a canonical form \cite{Henn:2013pwa}:
 \begin{align}
 &\nd f_I = \big( \nd \Omega \big)_{IK} f_K \, , \quad
 \big( \nd \Omega \big)_{IK} = \ep \sum_i \text{C}^{(i)}_{IK} ~ \nd \log \text{W}^{(i)}(\bm{s})  \, .  \label{eq:cde}
 \end{align} 
 In this expression, the $\big( \nd \Omega \big)_{IK}$ is a $\nd \log$-form proportional to 
 $\ep$. This property allows for the analytical solution of master integrals as multi-polylogarithm \cite{Chen:1977oja, Goncharov:1998kja}. This method  called Canonical Differential Equations (CDE)  has become a crucial analytic tool for evaluating Feynman integrals. The $\nd \log \text{W}^{(i)}(\bm{s})$ in \eqref{eq:cde} are referred to as symbol letters, and their complete set is known as the symbol alphabet. This set constrains the function space of analytic solutions and can be utilized for bootstrap techniques \cite{Gaiotto:2011dt, Dixon:2011pw, Dixon:2011nj, Brandhuber:2012vm, Caron-Huot:2019vjl}. Therefore, the properties of symbol letters, or $\nd \log \text{W}^{(i)}(\bm{s})$ in \eqref{eq:cde}, are important and have been studied extensively in \cite{Caron-Huot:2011dec, Golden:2013xva, Panzer:2014caa, Dennen:2015bet, Caron-Huot:2016owq, Mago:2020kmp, Abreu:2021vhb, Gong:2022erh, Yang:2022gko, He:2021non, He:2021eec, He:2022tph, He:2023qld, Arkani-Hamed:2017ahv, Abreu:2017enx, Abreu:2017mtm, Chen:2022fyw, Dlapa:2023cvx,Jiang:2024eaj}.

One way to 'appropriately select the master integrals' to obtain CDE is by constructing $\nd \log$ integrand as master integrands. Such construction is also related to leading singularities \cite{Henn:2013pwa, Chen:2020uyk, Chen:2022lzr, Chicherin:2018old, Bern:2014kca, Henn:2020lye, Dlapa:2021qsl}. In \cite{Chicherin:2018old,Chen:2020uyk, Chen:2022lzr}, it has been shown that Baikov representation  \cite{Baikov:1996iu} makes the construction of master integrands more convenient. Recently, \cite{Chen:2023kgw} showed how CDE emerge from $\nd \log$-form integrand of master integrals via intersection theory \cite{Mastrolia:2018uzb, Frellesvig:2019uqt}. In that work, one  only needs to use leading-order (LO) contribution and next-to-leading-order (NLO) contribution of intersection number. These contributions have universal formulas, and based on them, selection rules for CDE have been provided. However, in \cite{Chen:2023kgw}, the analysis requires the use of regulators to compute the intersection numbers, which are supposed to be set to zero at the end. The inclusion of these regulators introduces redundancies in both the calculations and the conclusions.

In this paper, we enhance the selection rules for CDE presented in \cite{Chen:2023kgw} in two aspects. Firstly, we present a new perspective on CDE by showing that the CDE matrix is the 1-$\mathrm{d}\log$-form coefficients obtained by projecting ($n$+1)-$\mathrm{d}\log$-forms onto $n$-$\mathrm{d}\log$-forms. This projection only requires the LO contribution to the intersection number, in contrast to the LO and NLO contributions used in \cite{Chen:2023kgw}. Secondly, we use the mathematical tool of twisted relative cohomology \cite{Caron-Huot:2021xqj,Caron-Huot:2021iev,Giroux:2022wav,De:2023xue} to avoid regulators. This approach helps us to    eliminate redundancy from the start. As a result, we obtain a more precise version of the CDE selection rules beyond \cite{Chen:2023kgw}.

\section{Notations of Intersection Theory and Relative Cohomology}
Intersection theory describes integrals of the form:
$ I[u,\varphi] \equiv \int  u \, \varphi \,, %\quad I_{\varphi_R} \equiv \int  u^{-1} \, \varphi_R \,,
$
where
\begin{align}
&\varphi \equiv \hat{\varphi}(\bm{z}) \bigwedge_j^n \nd z_j =\frac{Q(\bm{z})}{\left(\prod_k D_k^{a_k} \right) \left(\prod_i P_i^{b_i} \right)} \bigwedge_j^n \nd z_j \, , \quad  u=\prod_i \left[P_i(\bm{z})\right]^{\beta_i}  \, .
\label{eq:baikov}
\end{align}
Here $P_i$, $Q$ and $D_k$ are polynomials/monomials,  $a_k, b_j\in \mathbb{N}$ and $\beta_i\in \mathbb{C}$.
$I[u,\varphi]$ encompasses various integral representations, including Baikov representation and Feynman parameterization of Feynman integrals.

Then, IBP relations can be rewrite as 
\begin{equation}
\int\sum_i \partial_{z_i}(u \xi_i) \nd z_i = \int u (\varphi+\sum_i \nabla_i \xi_i)=0 \,, \, \quad \nabla_{i} = dz_i \wedge (\partial_{z_i} + \hat{\omega}_i) \, , \quad \hat{\omega}_i \equiv \partial_{z_i} \log(u) \,,
\end{equation}
where $u\xi_i$s are arbitrary integrands  belong to the integral family but in $(n-1)$-form.
Follow this expression, one can define the IBP-equivalence classes of cocycles \cite{Mastrolia:2018uzb, Frellesvig:2019uqt, Mizera:2019ose, Chestnov:2022xsy}:
\begin{equation}
\bra{\varphi_L} \equiv \varphi_L \sim \varphi_L + \sum_i \nabla_i \xi_i  \, .  \label{eq:braket}
\end{equation}
This leads to a cohomology group. Each integral $f_0=\int u \varphi_0$  corresponds to a vector in the cohomology group’s linear space \cite{Mastrolia:2018uzb, Frellesvig:2019uqt}. The master integrals $f_i=\int u \varphi_i$ form the basis of this space:
$
\bra{\varphi_0} = \sum_i c_i \bra{\varphi_i} .
$
The coefficients $c_i$ are determined by projecting rather than using IBP:
\begin{align}
&c_I=\sum_J \braket{\varphi_0|\varphi_J'}\left(\eta^{-1}\right)_{JI} \,, \quad \eta_{IJ}\equiv \braket{\varphi_I|\varphi_J'} \, , \label{eq:proj} \\
&\braket{\varphi_L|\varphi_R} = \sum_{\bm{p}} \Res_{\bm{z} = \bm{p}} \left(\psi_L \varphi_R\right) \, , \quad \nabla_1 \cdots \nabla_n \psi_L = \varphi_L \,. \label{eq:intnum}
\end{align}
Here the $\braket{\varphi_L|\varphi_R}$ denotes the intersection number, an IBP-invariant inner product.
In \eqref{eq:intnum},  $\bm{p}$ represents the isolated points where 
$n$ hypersurfaces in $\mathcal{B}=\{P_1=0,\infty,\cdots, D_1=0,\infty,\cdots\}$ intersect. Notice that $D_k$ in \eqref{eq:intnum} represent the denominators with integer powers in $u\varphi$.  In the context of Feynman integrals, such denominators typically arise from propagators. Direct computation of \eqref{eq:intnum} can be challenging with these denominators. One approach to manage this issue is to introduce regulators $\delta_k$ for such denominators in u: 
$
u=\prod_i \left[P_i(\bm{z})\right]^{\beta_i} \prod_k D_k^{\delta_k}
$
Afterward, the regulators are set to zero to obtain the final result. This method was used in  \cite{Chen:2023kgw}, but it led to redundancies.

In this paper, we use a technique known as relative cohomology to handle these denominators. While we do not delve into the mathematical details, the basic idea is as follows: we let the dual forms  $\varphi_R$ live in the maximal cut of the denominators and denote them as $\Delta_{\hat{R}}\varphi_R$. 
Practically, this is equivalent to applying the maximal cut of the sector of $\varphi_R$ to both sides before computing the intersection number.
\begin{align}
&\braket{\varphi_L|\Delta_{\hat{R}}\varphi_R}= \braket{\varphi_{L;\hat{R}}  |\varphi_{R;\hat{R}}} = \sum_{\bm{p}_{\hat{R}}}\Res_{\bm{p}_{\hat{R}}=0} \psi_{L;\hat{R}} \varphi_{R;\hat{R}}   \, , \quad \nabla_{1;\hat{R}} \cdots \nabla_{n;\hat{R}} \psi_{L;\hat{R}} = \varphi_{L;\hat{R}} \nn\\
& \varphi_{L;\hat{R}} = \Res_{\mathcal{B}_{\hat{R}}}  \left( \frac{u\varphi_L}{u_{\hat{R}}}\right)  \, , \quad  \varphi_{R;\hat{R}} = \Res_{\mathcal{B}_{\hat{R}}}  \left( \frac{\varphi_R u_{\hat{R}}}{u} \right)      \, ,  \nn\\
&u_{\hat{R}}=u|_{\mathcal{B}_{\hat{R}}}  \, , \quad  \mathcal{B}_{\hat{R}} = \{D_{R_1}=0,D_{R_2}=0,\cdots,D_{R_n}=0  \}  \, .  \label{eq:dualform}
\end{align}
Every symbol with the subscript $\hat{R}$ denotes the cut version of the original one. In \eqref{eq:dualform}, $D_{R_i}$  
 typically represent the propagators of the sector associated with $\varphi_R$. If $\mathcal{B}_{\hat{I}}  \subseteq \mathcal{B}_{\hat{J}} $, we say that $\mathcal{B}_{\hat{I}}$ is a subsector of $\mathcal{B}_{\hat{J}}$. 

To explicitly retain the information about multivariate poles in the calculations, we choose to analyze and compute the intersection number and the associated multivariate residue using a multivariate Laurent expansion. This expansion is applied after performing a factorization transformation corresponding to each pole \cite{Chestnov:2022xsy,Chen:2023kgw}:
\begin{align}
&\nT^{(\alpha)}: z_i \to f^{(\alpha)}_i(\bm{x}^{(\alpha)}) \,, \nn\\
&\nT^{(\alpha)}\left[u \right]\equiv   u(\nT^{(\alpha)}[\bm{z}]) = \bar{u}_\alpha(\bm{x}^{(\alpha)}) \prod_i \left[ x_i^{(\alpha)} - \rho_i^{(\alpha)} \right]^{\gamma_i^{(\alpha)}}  \, .
\label{eq:degenerate-1}
\end{align}
Only after this step one can legally apply a multivariate expansion to $\varphi$ and compute the intersection number:
\begin{align}
&\varphi= \sum_{\bm{b}} \varphi^{(\bm{b})}\,, \quad  \varphi^{(\bm{b})} = C^{(\bm{b})} \bigwedge_i  \left[ x_i^{(\alpha)} - \rho_i^{(\alpha)} \right]^{b_{i}} \nd x^{(\alpha)}_i \,,   \nn\\
&\braket{\varphi_L|\varphi_R}     = \sum_{\alpha} \Res_{\bm{\rho}^{(\alpha)}} \nT^{(\alpha)}\left[ \psi_L \varphi_R\right]  = \sum_{\alpha} \Res_{\bm{\rho}^{(\alpha)}} \left[ \left( \sum_{\bm{b}_L}  \nabla_1^{-1} \cdots \nabla_n^{-1} \varphi^{(\bm{b}_L)}_L  \right) \sum_{\bm{b}_R}\varphi^{(\bm{b}_R)}_R  \right] \, . \label{eq:intnumexpand}
\end{align}
Here, $\bm{b}=(b_1,\ldots,b_n)$ represents the powers in the Laurent expansion. When $b_{L,i}+b_{R,i} = -2$, we say it gives a LO contribution to intersection number. In such case, the contribution has a simple formula:
\begin{align}
&\Res_{\bm{\rho}^{(\alpha)}} \nT^{(\alpha)}\left[   \left( \nabla_1^{-1} \cdots \nabla_n^{-1}  \varphi^{(\bm{b}_L)}_L \right)\varphi^{(\bm{b}_R)}_R \right] =
\frac{C_L^{(\bm{b}_L)} C_R^{(\bm{b}_R)}}{\tilde{\bm{\gamma}}^{(\alpha)}} \,, \nn\\
&\tilde{\bm{\gamma}}\equiv \prod_i \tilde{\gamma}_i \,, \quad \tilde{\gamma}_i^{(\alpha)} = \gamma_i^{(\alpha)} - b_{R,i} - 1  \,, \ \ \  \bm{b}_L+\bm{b}_R=\bm{-2} \,  . \label{eq:intnumdlog}
\end{align}

\section{Selection Rules of CDE}
\subsection{$\nd \log$ projection and LO intersection number}
For clarity, we use $\nd$ to denote differentiation with respect to integration variables, and $\nds$ to denote differentiation with respect to one arbitrary parameter, such as kinetic parameters or masses. We define $\nD = \nd+\nds$.

To analyze and compute the CDE using \eqref{eq:proj} and \eqref{eq:dualform}, we start with the following equation:
\begin{align}
&\big( \nds \Omega \big)_{IK} = \braket{\dot{\varphi}_{I;\hat{J}}|\varphi_{J;\hat{J}}} \big( \eta^{-1} \big)_{JK} \, , \quad \eta_{IJ}\equiv \braket{\varphi_{I;\hat{J}}|\varphi_{J;\hat{J}}} \, .
\label{eq:intnumDE}
\end{align}
Here, $\varphi_I$ in both the original space and the dual space is the same $\nd \log$ integrand, constructed as $\varphi_I = \bigwedge_{j} \nd \log \nW^{(I)}_j(\bm{z})$.
The term $\dot{\varphi}_I$ in \eqref{eq:intnumDE} is defined as  $\nds f_I\equiv\int u \dot{\varphi}_I$. Therefore, we have:
\begin{align}
\nds f_I &= \int \nds \left( u \bigwedge_j \nd \log \nW^{(I)}_j(\bm{z}) \right) 
=\int u~\nD \log u \bigwedge_j \nD \log \nW^{(I)}_j  \, , \label{eq:Dlogvarphi}
\end{align}
 where $\dot{\varphi}_I$ is an ($n$+1)-$\nD \log$-form. To compute \eqref{eq:intnumDE},  we only need to compute LO contributions of the intersection number (after factorizing the pole), which are given by  \eqref{eq:intnumdlog}.

\subsection{Condition of $\braket{\dot{\varphi}_{I;\hat{J}}|\varphi_{J;\hat{J}}}\neq 0$}

Considering \eqref{eq:intnumdlog} and that $\varphi_I$ is a $\nd \log$-form, we can derive the condition under which $\braket{\dot{\varphi}_{I;\hat{J}}|\varphi_{J;\hat{J}}}\neq 0$:
\begin{align}
&\begin{cases}
\nd \log \text{-form} &: \  b_{I,i}\geq-1 \, , \ b_{J,i}\geq-1  \\
\dot{\varphi}_{I;\hat{J}} &:\  b_{\dot{I},j} = b_{I,j}-1  \\
\braket{\dot{\varphi}_{I;\hat{J}}|\varphi_{J;\hat{J}}}\neq 0 &: \ b_{\dot{I},i}+b_{J,i} \leq -2 
\end{cases} \nn\\
&\Rightarrow \braket{\dot{\varphi}_{I;\hat{J}}|\varphi_{J;\hat{J}}}\neq 0 :\  (-2\leq b_{I,j}+b_{J,j} \leq -1) \  \& \  (b_{I,i}=b_{J,i} = -1, i\neq j) \, ,
\end{align}
for specific one $j$. 
Here we omit the subscript "${;\hat{J}}$" of $b_{*,*}$ to simplify the expression, but it should be noted that all $b_{*,*}$ here correspond to the  powers after the maximal cut of the sector $\mathcal{B}_{\hat{J}}$.
We denote two cases for $\varphi_{I;\hat{J}}$ and $\varphi_{J;\hat{J}}$ sharing poles as follows:
\begin{itemize}
  \item Sharing \(n_{\hat{J}}\)-Variable Simple Pole (\(n_{\hat{J}}\)-SP): This occurs when \(b_{I,j} + b_{J,j} = -1\) for one specific \(j\) and \(b_{I,i} = b_{J,i} = -1\) for all \(i \neq j\) .
  \item Sharing \((n_{\hat{J}} - 1)\)-Variable Simple Pole (\((n_{\hat{J}} - 1)\)-SP): This occurs when \(b_{I,i} = b_{J,i} = -1\).
\end{itemize}

For the case in which they share an ($n_{\hat{J}}$-1)-SP,  $\nD \log \left( \nT^{(\alpha)} [u] \right)$ in \eqref{eq:Dlogvarphi} provides a pole of one remaining variable. The shared pole contributes
\begin{equation}
-\frac{\gamma^{(\alpha)}_{j;\hat{J}} }{\bm{\gamma}^{(\alpha)}_{\hat{J}} } \, \nds \int C_{I}^{(\bm{b}_{I} )}C_J^{(\bm{b}_J)} \, \nds \rho^{(\alpha)}_{j;{\hat{J}}} 
\label{eq:NMCSPdlog}
\end{equation}
to intersection number $\braket{\dot{\varphi}_{I;\hat{J}}|\varphi_{J;\hat{J}}}$.
It is a $\nds \log$-form as discussed in \cite{Chen:2023kgw}.

For the case in which they share an $n_{\hat{J}}$-SP, the shared pole contributes
\begin{equation}
\frac{C_I^{(\bm{-1})} C_J^{(\bm{-1})}}{\bm{\gamma}^{(\alpha)}_{\hat{J}}   } \, \nds \log \Big( \bar{u}_{\alpha;\hat{J}}(\bm{\rho}^{(\alpha)}_{\hat{J}} ) \Big)   \label{eq:CSPdlog}
\end{equation}
to intersection number $\braket{\dot{\varphi}_{I;\hat{J}}|\varphi_{J;\hat{J}}}$.

\subsection{Condition of $\big( \eta^{-1} \big)_{JK}\neq 0$}

Notice that $\eta_{JK}$ is non-zero only when $\varphi_{J;\hat{K}}$ and $\varphi_{K;\hat{K}}$ share at least an $n_{\hat{J}}$-SP and each shared $n_{\hat{J}}$-SP contributes $
C_J^{(\bm{-1})} C_K^{(\bm{-1})}/\bm{\gamma}^{(\alpha)}_{\hat{K}}  
$. Also, note that $\big( \eta^{-1} \big)_{JK} = (-1)^{J+K} |\eta^{(KJ)}| / |\eta|$, where $\eta^{(KJ)}$ is the adjoint matrix of $\eta$. Following a similar analysis as in \cite{Chen:2023kgw}, we need to modify the concept of "n-SP chain" from \cite{Chen:2023kgw}, which is unoriented, to a new concept of {\bf cut-$n$-SP} chain:
\begin{itemize}
\item If ${\varphi}_{I;\hat{J}}$ and $\varphi_{J;\hat{J}}$ share $n_{\hat{J}}$-SP,  we say $\varphi_I$ is cut-$n$-SP related to $\varphi_J$, and denoted as $\varphi_I \rightarrow \varphi_J$ or $\varphi_J \leftarrow \varphi_I$. Now, it is an orient relation, i.e., $\varphi_I \rightarrow \varphi_J$ does not imply $\varphi_J \rightarrow \varphi_I$. 

\item If $\varphi_I \rightarrow \varphi_J$, we also say that $\varphi_I$ is linked to $\varphi_J$ via a cut-$n$-SP chain. If $\varphi_I \rightarrow \varphi_J$ and $\varphi_J \rightarrow \varphi_K$, then we say $\varphi_I$ is linked to $\varphi_K$ via the cut-$n$-SP chain $\varphi_I \rightarrow \varphi_J \rightarrow \varphi_K$, we denote it as $\varphi_I \rightarrow\rightarrow \varphi_J$. Similar understanding for more forms $\varphi$.  
\end{itemize}
Then, \(\big( \eta^{-1} \big)_{JK}\) could be non-zero only when \(\varphi_J \rightarrow\rightarrow \varphi_K\).

\subsection{Conclusion}
With the above discussions,  we have CDE selection rules: 
If all master integrals have $\nd \log$-form $\varphi_I$s, the differential equations $\big( \nds \Omega \big)_{IK}$  is $\nd \log$-form and given by the simple formula of  LO contribution of intersection number. When all $\beta_i$ in eq.\eqref{eq:baikov} are proportional to $\ep$, $\big( \nds \Omega \big)_{IK}$ is reduced to be proportional to $\ep$, thus is canonical form. $\big( \nds \Omega \big)_{IK}$ could be non-zero only when there exist a $\varphi_J$ that  $\dot{\varphi}_{I;\hat{J}}$ and $\varphi_{J;\hat{J}}$ share $n_{\hat{J}}$-SP or ($n_{\hat{J}}-1$)-SP, and $\varphi_J \rightarrow \rightarrow \varphi_K$.

\section{Univariate Example: Comparing Regulator Method with Relative Cohomology}
Consider $u=z^\delta (z-c_1)^\ep(z-c_2)^\ep$, where  $\delta$ 
 is regulator. We construct the $\nd \log$ basis as follow:
\begin{align}
\varphi_1 = \frac{\nd z}{z} =\nd \log z \, ,  \quad \varphi_2 = \frac{\nd z}{z-c_1} - \frac{\nd z}{z-c_2}  = \nd \log \left( \frac{z-c_1}{z-c_2}\right)  \, .
\end{align}

\subsection{Regulator method}
If we compute the differential equations of the basis using the regulator method, we obtain:
\begin{align}
&\omega=\nd \log u = \left( \frac{\delta}{z} + \frac{\ep}{z-c_1}+ \frac{\ep}{z-c_2} \right) \nd z \, . \nn\\
&\mathcal{B}=\{z=0,z-c_1=0,z-c_2=0,z=\infty\}\,,\quad \{\bm{p}\}=\{0,c_1,c_2,\infty\} \, . \nn\\
&\gamma_1=\delta\, , \ \ \gamma_2=\ep\, , \ \ \gamma_3=\ep\, , \ \ \gamma_4=-2\ep-\delta \, .
\end{align}
Then, $\eta_{11}$ receives contributions from shared 1-SP at $p_1=0$ and $p_4=\infty$ of $\varphi_1$ itself, so it equals $1/\gamma_1+1/\gamma_4$. A similar analysis can be applied to other \(\eta_{JK}\), giving:
\begin{align}
&\eta=\braket{\varphi_I|\varphi_J}=\left(
\begin{array}{cc}
\frac{1}{\gamma_1}+\frac{1}{\gamma_4} & 0 \\
0 &  \frac{1}{\gamma_2}+\frac{1}{\gamma_3}  \\
\end{array}
\right)=\left(
\begin{array}{cc}
\frac{1}{\delta }-\frac{1}{\delta +2 \epsilon } & 0 \\
0 & \frac{2}{\epsilon } \\
\end{array}
\right)  \, , \nn\\
&\eta^{-1}=\left(
\begin{array}{cc}
\delta(\delta+2\ep)/(2 \epsilon)  & 0 \\
0 & \epsilon/2 \\
\end{array}
\right)  \, .
\end{align}
For $\braket{\dot{\varphi}_I|\varphi_J}$, we have $\dot{\varphi}_I$
\begin{align}
&\dot{\varphi}_1 = \nD \left(\log \bar{u}_1(z) + \delta \log z \right) \wedge \nD \log z  = \nD \log \bar{u}_1(z) \wedge \nD \log z  \, , \nn\\
&\dot{\varphi}_2 =  \nD \log \bar{u}_2(z) \wedge \nD \log (z-c_1) - \nD \log \bar{u}_3(z) \wedge \nD \log (z-c_2)   \nn\\
&\quad = \delta ~ \nD \log z \wedge \varphi_2 + 2 \ep ~ \nD \log (z-c_2) \wedge \nD \log (z-c_1)  \, , \nn\\
&\bar{u}_1(z) = (z-c_1)^\ep(z-c_2)^\ep \, , \quad  \bar{u}_2(z) = z^\delta(z-c_2)^\ep \, , \quad  \bar{u}_3(z) = z^\delta(z-c_1)^\ep \, .
\end{align}
Let's calculate $\braket{\dot{\varphi}_1|\varphi_1}$ as an example. For the contribution from the shared 1-SP $p_1=0$, we use \eqref{eq:CSPdlog} to get
\begin{align}
\frac{1}{\delta} \nds \log \left((z-c_1)^\ep(z-c_2)^\ep|_{z=0} \right) = \frac{\ep}{\delta} \nds \log(c_1c_2)  \,,
\end{align}
For the shared 1-SP at $p_4=\infty$:
\begin{align}
\frac{1}{-\delta-2\ep} \nds \log \left((1-c_1 t)^\ep(1-c_2 t)^\ep|_{t=0} \right) = \frac{\ep}{-\delta-2\ep} \nds \log 1 = 0 \,,
\end{align}
where $t=1/z$. Thus, this term does not contribute. Continuing the calculation, we find:
\begin{align}
&\Big( \braket{\dot{\varphi}_I|\varphi_J} \Big) = \nds \left(
\begin{array}{cc}
\frac{\epsilon }{\delta } \log \left(c_1c_2\right) & \log
\left(c_2/c_1\right) \\
\log \left(c_2/c_1\right) & \frac{\delta }{\epsilon } \log \left(c_1c_2\right) + 4\log(c_1-c_2) \\
\end{array}
\right)  \, .  \nn\\
&\nds \Omega^{\text{reg}} (c_i;\ep,\delta) = \nds \left(
\begin{array}{cc}
\frac{\delta +2 \ep}{2} \log \left(c_1 c_2\right)& \frac{\epsilon}{2}   \log \left(\frac{c_2}{c_1}\right) \\
\frac{\delta   (\delta +2 \epsilon )}{2 \epsilon } \log \left(\frac{c_2}{c_1}\right) 
& \frac{\delta}{2}   \log \left(c_1 c_2\right)+2 \epsilon  \log \left(c_1-c_2\right) \\
\end{array}
\right)   \, . \label{eq:1varDEreg}
\end{align}
It is redundant and $\nds \Omega^{\text{reg}}_{21}$ vanishes when $\delta=0$, but the selection rules get from the regulator method in \cite{Chen:2023kgw} do not  indicate this.

\subsection{Relative cohomology method}
If we calculate $\Omega$ using relative cohomology method, we have 
\begin{align}
&u= (z-c_1)^\ep(z-c_2)^\ep
\, , \ \  D_1=z   \nn\\
&\omega=\nd \log u = \left( \frac{\ep}{z-c_1}+ \frac{\ep}{z-c_2} \right) \nd z \, . \quad
\{\bm{p}\}=\{0,c_1,c_2,\infty\} \, .
\end{align}
Calculation of $\braket{\varphi_I|\Delta_{\hat{1}}\varphi_1}$ is simplified as we apply the cut of $\mathcal{B}_{\hat{1}}$:
\begin{align}
&\braket{\varphi_1|\Delta_{\hat{1}}\varphi_1}=\braket{\varphi_{1;\hat{1}}|\varphi_{1;\hat{1}}} = \braket{1|1}=1  \nn\\
&\braket{\varphi_2|\Delta_{\hat{1}}\varphi_1}=\braket{\varphi_{2;\hat{1}}|\varphi_{1;\hat{1}}} = \braket{0|1}=0 \nn\\
&\braket{\dot{\varphi}_1|\Delta_{\hat{1}}\varphi_1}=\braket{\dot{\varphi}_{1;\hat{1}}|\varphi_{1;\hat{1}}} = \braket{\nds\log u|_{z=0}|1}= \ep \nds \log (c_1 c_2) \nn\\
&~\dot{\varphi}_2= \nD \log u \wedge \nD \log \left(z-c_1\right) - \nD \log u \wedge \nD \log \left(z-c_2\right) \nn\\
& \quad ~ = 2 \ep ~  \nD \log \left(z-c_2\right) \wedge \nD \log \left(z-c_1\right) \, , \nn\\
&\braket{\dot{\varphi}_2|\Delta_{\hat{1}}\varphi_1}=\braket{\dot{\varphi}_{2;\hat{1}}|\varphi_{1;\hat{1}}} = \braket{0|1}= 0 \, .
\end{align}
Thus, we have
\begin{align}
&\eta=\braket{ \varphi_{I;\hat{J}}|\varphi_{J;\hat{J}}} =\left(
\begin{array}{cc}
1 & 0 \\
0 & 2/\epsilon  \\
\end{array}
\right)  \, , \quad  \eta^{-1}=\left(
\begin{array}{cc}
1  & 0 \\
0 & \epsilon/2 \\
\end{array}
\right)  \, , \nn\\
&\Big( \braket{\dot{\varphi}_{I;\hat{J}}|\varphi_{J;\hat{J}}} \Big) = \nds \left(
\begin{array}{cc}
\ep \log \left(c_1 c_2\right) 
& \ep \log \left(c_2/c_1\right) \\
0 
& 4 \log
\left(c_1-c_2\right) \\
\end{array}
\right)  \, \nn\\
&\nds \Omega =\Big( \braket{\dot{\varphi}_{I;\hat{J}}|\varphi_{J;\hat{J}}} \Big).\eta^{-1}= \ep~  \nds \left(
\begin{array}{cc}
\log \left(c_1 c_2\right) & \frac{1}{2} \log \left(c_2/c_1\right) \\
0 & 2  \log
\left(c_1-c_2\right) \\
\end{array}
\right)  \, .
\end{align}
We have $\nds \Omega=\nds \Omega^{\text{reg}} (c_i;\ep,0)$. This demonstrates that the relative cohomology method simplifies the computation, avoids redundancy from regulators, and provides a more precise version of the selection rules for CDE. In particular, it correctly shows that $\Omega_{21}=0$ in this case. 

\section{Summary and Outlook}
We improve the selection rules for CDE by providing a perspective from the $\nd \log$ projection and avoiding redundancy through the relative cohomology method. The mathematical structure we have developed can benefit various aspects of perturbative quantum field theory, including reduction (since CDEs are themselves reduction relations), symbolic bootstrap, and the analytic evaluation of Feynman integrals.

\section*{Acknowledgements}
We thank Andrzej Pokraka for valuable discussions. This work is supported by Chinese NSF funding under Grant No.11935013, No.11947301, No.12047502 (Peng Huanwu Center), No.12247120, No.12247103, NSAF grant No.U2230402, and China Postdoctoral Science Foundation No.2022M720386.

\vspace{-1mm}
\setlength{\bibsep}{2pt}
\bibliographystyle{JHEP}
\bibliography{ref}

\end{document}